\documentclass[aps, prl, showpacs, nofootinbib, twocolumn]{revtex4}
\usepackage{amssymb,amsmath,microtype}
\usepackage{graphicx}
\usepackage[normalem]{ulem}

\usepackage{color}

\begin{document}


\title{Hearing the Echoes of Electroweak Baryogenesis with Gravitational Wave Detectors}

\author{Fa Peng Huang$^{1}$}
\author{Youping Wan$^{1}$}
\author{Dong-Gang Wang$^{2}$}
\author{Yi-Fu Cai$^{2}$}
\author{Xinmin Zhang$^{1}$}

\affiliation{$^1$Theoretical Physics Division, Institute of High Energy Physics, Chinese Academy of Sciences, P.O.Box 918-4, Beijing 100049, P.R.China}
\affiliation{$^2$CAS Key Laboratory for Researches in Galaxies and Cosmology, Department of Astronomy, University of Science and Technology of China, Chinese Academy of Sciences, Hefei, Anhui 230026, China}

\begin{abstract}
We report on the first joint analysis of observational signatures from the electroweak baryogenesis in both gravitational wave (GW) detectors and particle colliders.
With an effective extension of the Higgs sector in terms of the dimension-6 operators, we derive a strong first-order phase transition in associated with a sizable CP violation to realize a successful electroweak baryogenesis.
We calculate the GW spectrum resulting from the bubble nucleation, plasma transportation, and magnetohydrodynamic turbulence of this process that occurred after the big bang, and find that it yields GW signals
testable in Evolved Laser Interferometer Space Antenna, Deci-hertz Interferometer Gravitational wave Observatory and Big Bang Observer.
We further identify collider signals from the same mechanism that are observable at the planning Circular Electron Positron Collider.
Our analysis bridges astrophysics/cosmology with particle physics by providing significant motivation for searches for GW events
peaking at the $(10^{-4}, 1)$ Hz range, which are
associated with signals at colliders, and highlights the possibility of an interdisciplinary observational window into baryogenesis. 
The technique applied in analyzing early universe phase transitions may enlighten the study of phase transitions in applied science.
\end{abstract}

\pacs{04.30.-w, 12.60.-i, 95.55.-n}

\maketitle

{\it Introduction.---}
The Advanced Laser Interferometer Gravitational Wave Observatory (aLIGO) recently reported the first direct detection of gravitational waves (GW) from the coalescence of black hole binary~\cite{Abbott:2016blz}. This breakthrough is expected to initiate a novel probe of cosmology, the nature of gravity as well as the fundamental physics.

The Universe experienced phase transitions after the big bang. If they were of first order, then one major consequence would be an existence of echoes of GW in early Universe~\cite{Witten:1984rs, Hogan:1984hx, Turner:1990rc, Schwaller:2015tja}. Among them, the electroweak phase transition (EWPT) is one significant target of particle physics following the discovery of the Higgs boson~\cite{Aad:2012tfa} since it is closely related to the new physics beyond the standard model (SM)~\cite{CEPC-SPPCStudyGroup:2015csa, Arkani-Hamed:2015vfh}.
At present, our knowledge about the nature of the Higgs field remains scarce since very limited information can be learned from current particle colliders. Without new observational windows, one cannot distinguish the tree-level Higgs potential to be the SM form
or others involving high-dimension operators.
Theoretically, the Higgs scenario including a sextic term can yield a strong first-order phase transition (SFOPT) for electroweak (EW) baryogenesis~\cite{Zhang:1992fs, Zhang:1993vh, Grzadkowski:2003sd, Grojean:2006bp, Delaunay:2007wb, Huber:2007vva, Das:2009ue, Espinosa:2008kw, Jarvinen:2009mh, Kakizaki:2015wua, Huang:2015bta, Huang:2015izx}. Therefore, it is essential to properly characterize the predicted GW spectra from these transitions.


In this Letter we report on the first joint analysis of observational signatures from the EW baryogenesis, which could have occurred at the early Universe, in both GW and collider experiments. Considering an effective field theory (EFT) extension of the Higgs Lagrangian with a sextic term, which represents for new physics beyond the SM, a SFOPT can be realized to generate GW relics. We numerically calculate their energy spectrum under a series of cosmological effects including the bubble nucleation, plasma transportation, and magnetohydrodynamic (MHD) turbulence. Our results show that the corresponding GW signals are lower than the sensitivity of aLIGO \cite{TheLIGOScientific:2014jea} and Virgo~\cite{TheVirgo:2014hva}, but can be testable in other surveys like Evolved Laser Interferometer Space Antenna (eLISA)~\cite{Seoane:2013qna}, Deci-hertz Interferometer Gravitational wave Observatory (DECIGO)~\cite{Seto:2001qf} and Big Bang Observer (BBO)~\cite{Corbin:2005ny}. The same mechanism generates a nontrivial trilinear Higgs coupling that could be examined at the lepton collider of new generation, the Circular Electron Positron Collider (CEPC)~\cite{Huang:2015izx}. Our analysis reveals an interesting phenomenon that each signal of the Higgs-induced EW baryogenesis at the collider is associated with an unique pattern of the GW spectrum for astronomical survey.

{\it An effective theory of EW baryogenesis and Collider signals.---}
Instead of investigating the EWPT/baryogenesis in a UV-complete theory, which is difficult to make experimental predictions from unknown model parameters, we take a bottom-up approach to explain the baryon asymmetry of the universe and study the possible collider and GW signals. Then, utilizing the EFT approach, one may write the effective Lagrangian of the Higgs doublet $\phi$ as follows,
 $\delta\mathcal{L} = -x_{u}^{ij} \frac{\phi^\dagger_{}\phi}{\Lambda^2_{}}\bar{q}^{}_{Li} \tilde{\phi} u^{}_{Rj} +\textrm{h.c.} -\frac{\kappa}{\Lambda^2_{}}(\phi^\dagger_{}\phi)^3_{}$,
where $\tilde{\phi} \equiv i \tau_2 \phi^*$, $q_L^{}$ and $u_R^{}$ are respectively the left-handed quarks and the right-handed up-type quarks. Moreover, $\kappa$ and $\Lambda$ respectively correspond to a coupling parameter and a cutoff scale~\cite{Zhang:1992fs, Zhang:1993vh}.
These effective operators could come from renormalizable extensions of the SM, namely, models with vector-like quarks and a triplet Higgs~\cite{Huang:2015izx} or with additional scalar fields~\cite{CEPC-SPPCStudyGroup:2015csa,Arkani-Hamed:2015vfh}. Note that, the last operator is able to realize a SFOPT and the first two can induce a sizable CP violation.

To investigate the EWPT, it is convenient to work with the unitary gauge $\phi=h/\sqrt{2}$. Accordingly, the tree-level Higgs potential becomes:
\begin{equation}\label{v0}
V_{\rm tree}(h) = \frac{1}{2}\mu^2 h^2 + \frac{\lambda}{4} h^4 + \frac{\kappa}{8\Lambda^2} h^6,
\end{equation}
and the one-loop finite-temperature effective potential can be written as
 $V_\mathrm{eff}(h,T) = V_\text{tree}(h) + V_1^{T=0}(h) + \Delta V_1^{T\neq 0}(h,T)$,
with $V_1^{T=0}(h)$ being the one-loop Coleman-Weinberg potential at $T=0$, and $\Delta V_1^{T\neq 0}(h)$ the thermal contribution with the daisy resummation~\cite{Quiros:1999jp}. In this type of model the dominant contribution for the EWPT is from the tree-level barrier, and hence, the effective potential with finite temperature effects approximately takes
$V_{\rm eff}(h, T) \approx \frac{\kappa}{8\Lambda^2} h^6+ \frac{\lambda}{4} h^4 + \frac{1}{2} ( \mu^2 + c \, T^2 ) h^2,$
with
$c=\frac{1}{16}(-12\frac{\kappa v^2}{\Lambda^2}+g'^2+3 g^2+4 y_t^2+4\frac{m_h^2}{v^2}),$
where the coefficients $g'$ and $g$ are the $U(1)_Y$  and $SU(2)_L$ gauge couplings, respectively, and $y_t$ is the top quark Yukawa coupling in the SM.
From the standard analysis of the EW baryogenesis, the critical temperature $T_c >0$ and the washout factor $v(T_c)/T_c >1$ give the constraints on the
cutoff scale $\Lambda_{\rm min}<\Lambda<\Lambda_{\rm max}$, with $\Lambda_{\rm max} \equiv \sqrt{3\kappa} v^2 / m_h$ and $\Lambda_{\rm min} \equiv \Lambda_{\rm max} / \sqrt{3} =\sqrt{\kappa} v^2 / m_h$.
To fix the observed Higgs mass $m_h=125~\rm GeV$ and the vacuum expectation value $v$, the parameters $\lambda$ and $\mu^2$ satisfy the relations: $\lambda = \lambda_{\rm SM} \big( 1 -\frac{\Lambda_{\rm max}^2}{\Lambda^2} \big)$ and $\mu^2 = \mu^2_{\rm SM}  \big( 1 -\frac{\Lambda_{\rm max}^2}{2 \Lambda^2} \big)$,
with $\Lambda_{\rm max} \equiv \sqrt{3\kappa} v^2 / m_h$.
In addition, the perturbativity requires that $\kappa < 4 \pi$. If one chooses a larger $\kappa$, however, a larger bound for $\Lambda_{\rm max}$ may be achieved. For $m_h = 125~{\rm GeV}$, there is $480~\rm GeV<\Lambda/\sqrt{\kappa}<840~\rm GeV$, as required by the SFOPT.


A novel consequence of this effective theory is that the requirement of the SFOPT can lead to an obvious modification of the trilinear Higgs coupling as
$\mathcal{L}_{hhh}= -\frac{1}{6} (1+ \delta_h) A_{h} h^3,$
with $A_{h}=3 m_h^2/v$ being the trilinear Higgs coupling in the SM and $\delta_h=2 \Lambda_{\min}^2/\Lambda^2$. In our model $\delta_h$ varies from $2/3$ to $2$. It turns out that one can test the EW baryogenesis by probing the deviation of the trilinear Higgs coupling at colliders. For the Large Hadron Collider (LHC), such a deviation leads to different invariant mass distribution from the SM one. However, due to the challenge of suppressing the large backgrounds at hadron colliders, the trilinear Higgs coupling is difficult to be pinned down at the 14 TeV LHC. Interestingly, for lepton colliders, namely, the International Linear Collider (ILC) and CEPC, the trilinear Higgs coupling could be measured precisely. In particular, at the CEPC with $\sqrt{s}=240~\rm GeV$, the one-loop contribution to $hZ$ cross section ($\sigma_{hZ}$) beyond the SM will be dominated by the modified trilinear Higgs coupling~\cite{Huang:2015izx}. Therefore, a deviation of $\sigma_{hZ}$, which is defined as $\delta_{\sigma_{hZ}} \equiv {\sigma_{hZ}}/{\sigma_{hZ}^\mathrm{SM}} -1$, can be induced and it is approximately proportional to $\delta_h$ as $\delta_{\sigma_{hZ}}\simeq 1.6\% ~\delta_h$ at $\sqrt{s}=240~\rm GeV$.
Thus, for $\kappa=1$, one gets
$\delta_{\sigma_{hZ}} \simeq {7514.17~\rm GeV^2}/{\Lambda^2}.$
For the CEPC with an integrated luminosity of $10~\mathrm{ab}^{-1}$, the precision of $\sigma_{hZ}$ could be $0.4\%$~\cite{Gomez-Ceballos:2013zzn}, which corresponds to $|\delta_h| \sim 25\%$. In our scenario, $\delta_h \in (2/3,2)$, and hence, the associated signals could be observable at the CEPC. More connections between the Higgs trilinear coupling can be found in \cite{Grojean:2004xa, Noble:2007kk}.

{\it GW signals of EW baryogenesis.---}
For the Higgs potential responsible for EW baryogenesis, there exists a potential barrier between the metastable false vacuum and the true one. If the EWPT is strong enough, vacuum bubbles are nucleated via quantum tunneling. The temperature goes down along with the cosmic expansion, and the nucleation probability of one bubble per one horizon volume becomes larger and larger. The EWPT completes when the probability is of $\mathcal{O}(1)$ at the transition temperature, i.e., $\Gamma(T_{\ast}) \simeq H_{\ast}^4$, and then, we obtain
$S_3(T_{\ast})/T_{\ast} =4\ln (T_{\ast}/100 \mbox{GeV})+137$,
where
$ S_3 \equiv \int d^3r [\frac{1}{2}  (\vec{\nabla}h)^2+ V_{\rm eff}(h, T)]$
is the three dimensional Euclidean action.

The properties of the EWPT and of the bubbles are determined by two key parameters $\alpha$ and $\beta$. Note that, $\alpha$ is defined by $\alpha \equiv \frac{\epsilon(T_{\ast})}{\rho_{\rm rad}(T_{\ast})}$ at the transition temperature $T_{\ast}$, which depicts the ratio of the false vacuum energy density $\epsilon(T)$ (the latent heat where $\epsilon(T_{\ast}) = [T \frac{dV_{\rm eff}^{\rm min}}{dT} -V_{\rm eff}^{\rm min}(T) ]|_{T=T_{\ast}}$) to the plasma thermal energy density $\rho_{\rm rad}(T)$ (which is equal to $\frac{\pi^2}{30} g_{*}(T)T^4$) in the symmetric phase. Moreover, one has $\beta \equiv -\frac{d S_E}{d t} |_{t=t_{\ast}} \simeq\frac{1}{\Gamma}\frac{d \Gamma}{d t} |_{t=t_{\ast}}$, where $S_E(T)\simeq S_3(T)/T$, and $\Gamma=\Gamma_0(T)\exp[-S_E(T)]$ represents the variation of the bubble nucleation rate with $\Gamma_0(T)\propto T^4$. The parameter $\alpha$ gives a measure of the strength of the EWPT, namely, a larger value for $\alpha$ corresponds to a stronger EWPT. Furthermore, $\beta^{-1}$ corresponds to the typical time scale of the EWPT and its product with the bubble wall velocity $\beta^{-1}v_b(\alpha)$ represents the size of the bubble.  These derived parameters for different cutoff scales $\Lambda$ are listed in Table \ref{alphabeta}.

\begin{table}
    \begin{tabular}{|c|c|c|c|}
    \hline
   $\Lambda$          &~  $T_*$          ~  & ~$\alpha$~ & ~$\beta/H_*$~  \\ \hline
  ~590~GeV  ~       &~ 40.62~GeV   ~  & ~0.66 ~      & ~138.1~ \\ \hline
  ~600~GeV  ~       &~ 51.94~GeV   ~   & ~0.29 ~     & ~ 346.1~\\ \hline
  ~650~GeV  ~       &~ 75.42~GeV   ~   & ~0.09 ~     & ~ 1696.1~\\ \hline
  ~700~GeV  ~       &~ 87.60~GeV   ~   & ~0.05 ~     & ~ 7980.7~ \\ \hline
  ~750~GeV  ~       &~ 96.08~GeV   ~   & ~0.03 ~      & ~ 26486.2~ \\ \hline
    \end{tabular}
    \caption {The derived parameters of EWPT for different cutoff scales $\Lambda$.}\label{alphabeta}
\end{table}

It is known that there exist three major sources for producing GW during SFOPT, which respectively are collisions of the vacuum bubbles~\cite{Kamionkowski:1993fg}, sound waves~\cite{Hindmarsh:2013xza} and MHD turbulence~\cite{Kosowsky:2001xp, Caprini:2009yp} in the plasma after collisions.
The peak frequency produced by bubble collisions at the time of phase transition is given by~\cite{Huber:2008hg}: $f_{\rm co}^\ast=0.62\beta/(1.8-0.1v_{b} +v_{b}^{2})$.
Considering the adiabatic expansion from the radiation dominated stage to the present universe, we get the ratio of scale factors at EWPT and today
\begin{align}
\frac{a_\ast}{a_0}= 1.65 \times 10^{-5} \mbox{Hz}\times\frac{1}{H_{\ast}} \Big( \frac{T_{\ast}}{100 \mbox{GeV}} \Big) \Big( \frac{g^t_\ast}{100} \Big)^{1/6}, \nonumber
\end{align}
where $g^t_\ast$ is the total number of degrees of freedom at $T_{\ast}$. As a result, the peak frequency becomes $f_{\rm co}= f_{\rm co}^\ast a_\ast/a_0$ today, and the corresponding GW intensity is calculated as \cite{Huber:2008hg}
\begin{align}
 \Omega_{\rm co} (f) h^2 \simeq
 &1.67\times 10^{-5} \Big( \frac{H_{\ast}}{\beta} \Big)^2 \Big( \frac{\varepsilon \alpha}{1+\alpha} \Big)^2 \Big( \frac{100}{g^t_\ast} \Big)^{\frac{1}{3}} \nonumber \\
 &\times \Big( \frac{0.11v_b^3}{0.42+v_b^3} \Big) \Big[ \frac{3.8(f/f_{\rm co})^{2.8}}{1+2.8(f/f_{\rm co})^{3.8}} \Big]. \nonumber
\end{align}
The coefficient $\varepsilon$ (which characterizes the fraction of the latent heat that is transformed to the fluid kinetic energy) and the bubble wall velocity $v_b$ are functions of $\alpha$\cite{Kamionkowski:1993fg}. For this part of contribution, in the low frequency regime the spectrum $\Omega_{\rm co} h^2$ increases as $f^{2.8}$, but in the high frequency regime it decreases as $f^{-1}$~.

The GW signals due to the sound wave effects yield a peak frequency at about $f_{\rm sw}^\ast=2\beta/{\sqrt{3}v_b}$ \cite{Hindmarsh:2013xza, Caprini:2015zlo}, and similarly its current value takes $f_{\rm sw}= f_{\rm sw}^\ast a_\ast/a_0$.
In this case, the GW intensity is expressed as \cite{Hindmarsh:2013xza, Caprini:2015zlo}
\begin{align}
\Omega_{\rm sw} (f) h^2\simeq \nonumber
& 2.65\times 10^{-6}\Big(\frac{H_{\ast}}{\beta}\Big) \Big(\frac{\varepsilon_{\nu} \alpha}{1+\alpha}\Big)^2
\Big(\frac{100}{g^t_\ast}\Big)^{\frac{1}{3}}v_b\\
&\times\Big[\frac{7(f/f_{\rm sw})^{6/7}}{4+3(f/f_{\rm sw})^2}\Big]^{7/2}, \nonumber
\end{align}
in which the factor $\varepsilon_{\nu}$ represents the fraction of latent heat that is transformed into bulk motion of the fluid. Note that, $\varepsilon_{\nu}\simeq \alpha \left(0.73+0.083\sqrt{\alpha}+\alpha\right)^{-1}$ for relativistic bubbles~\cite{Espinosa:2010hh}. One observes that, the GW spectrum arisen from the sound wave effects, $\Omega_{\rm co} h^2$, evolves as $f^{3}$ in the low frequency regime but then becomes $f^{-4}$ in the high frequency regime.

The GW signals produced by the MHD turbulence in the plasma have a peak frequency at about $f_{\rm tu}^\ast=3.5\beta/2v_b$ \cite{Caprini:2015zlo}, which determines the present one as $f_{\rm tu}= f_{\rm tu}^\ast a_\ast/a_0$ after redshifting.
This part of the GW intensity is formulated by \cite{Caprini:2009yp, Binetruy:2012ze}
\begin{align}
\Omega_{\rm tu} (f) h^2\simeq \nonumber
& 3.35\times 10^{-4}\Big(\frac{H_{\ast}}{\beta}\Big) \Big(\frac{\varepsilon_{\rm tu} \alpha}{1+\alpha}\Big)^{3/2}
\Big(\frac{100}{g^t_\ast}\Big)^{\frac{1}{3}}v_b\\
&\times\frac{(f/f_{\rm tu})^3}{(1+f/f_{\rm tu})^{11/3}(1+8\pi fa_0/(a_\ast H_\ast))}, \nonumber
\end{align}
where $\varepsilon_{\rm tu}\simeq 0.1\varepsilon_{\nu}$. The GW spectrum contributed by the MHD turbulence, $\Omega_{\rm tu} h^2$, is approximately proportional to $f^{3}$ in the low frequency regime but takes $f^{-2/3}$ in the high frequency regime.
Accordingly, it is interesting to notice that the EWPT has predicted particular patterns of the intensity spectrum in terms of the above three parts, which may be key signatures in GW surveys.
It is worth noting that, however, the bubble wall runs away if $\Lambda$ becomes smaller than 590 GeV~\cite{Bodeker:2009qy,Huber:2013kj}.



\begin{figure}
\begin{center}
\includegraphics[scale=0.45]{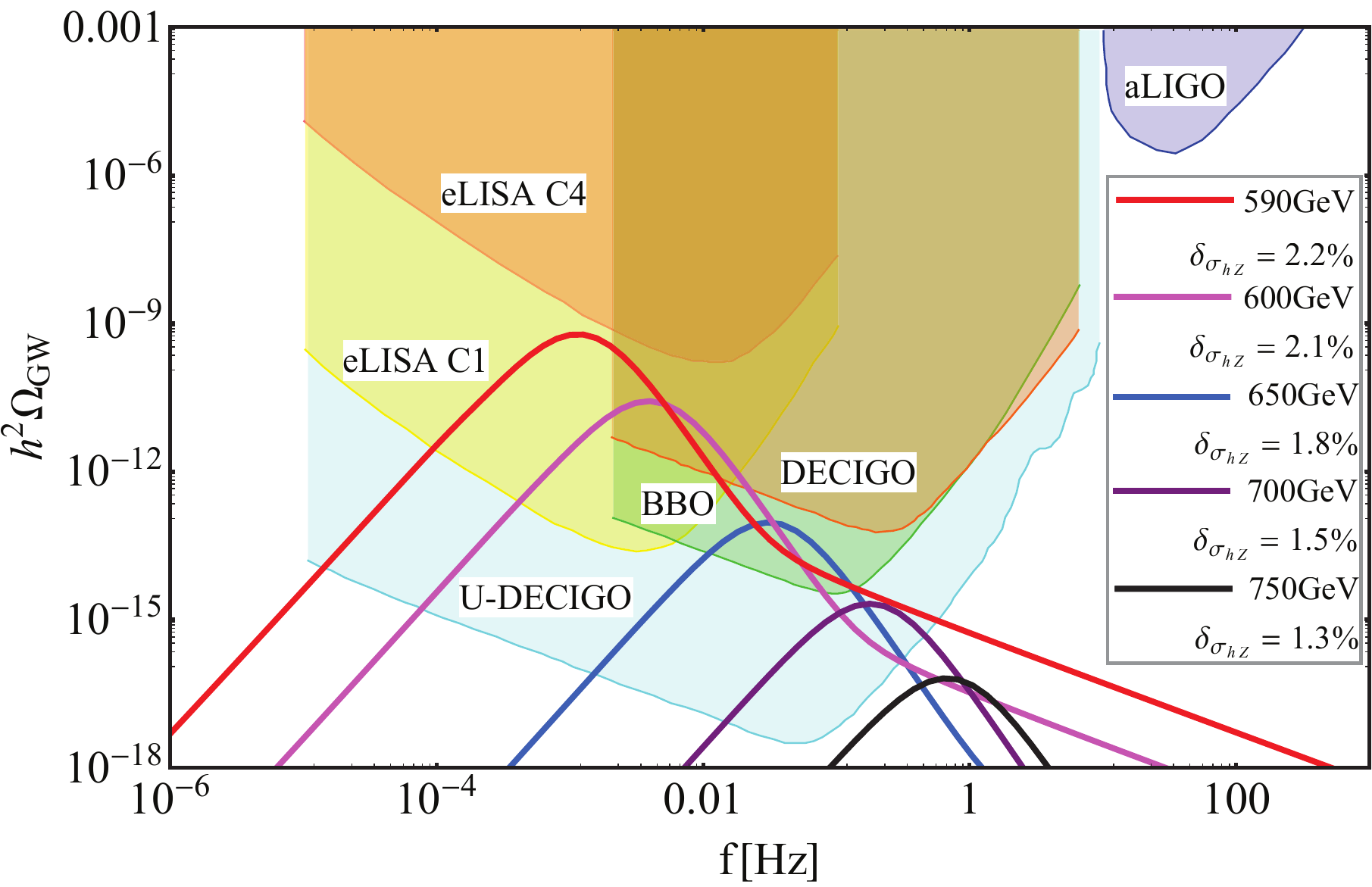}
\caption{The GW spectra $h^2\Omega_{GW}$ and the associated collider signals $\delta_{\sigma_{hZ}}$ for different cutoff scales $\Lambda$ (590~GeV, 600~GeV, 650~GeV,~700~GeV and 750 GeV) with $\kappa=1$. The colored regions correspond to the expected sensitivities of GW interferometers aLIGO, eLISA, BBO, DECIGO, and U-DECIGO. The red line depicts the GW spectrum for $\Lambda = 590~{\rm GeV}$, which is related to a collider signal of $\delta_{\sigma_{hZ}} \simeq 2.2\%$ at the CEPC. The magenta, blue, purple and black lines are the cases for 600 GeV, 650 GeV, 700 GeV and 750 GeV, respectively.
}
\label{gw}
\end{center}
\end{figure}

\begin{figure}
\begin{center}
\includegraphics[scale=0.3]{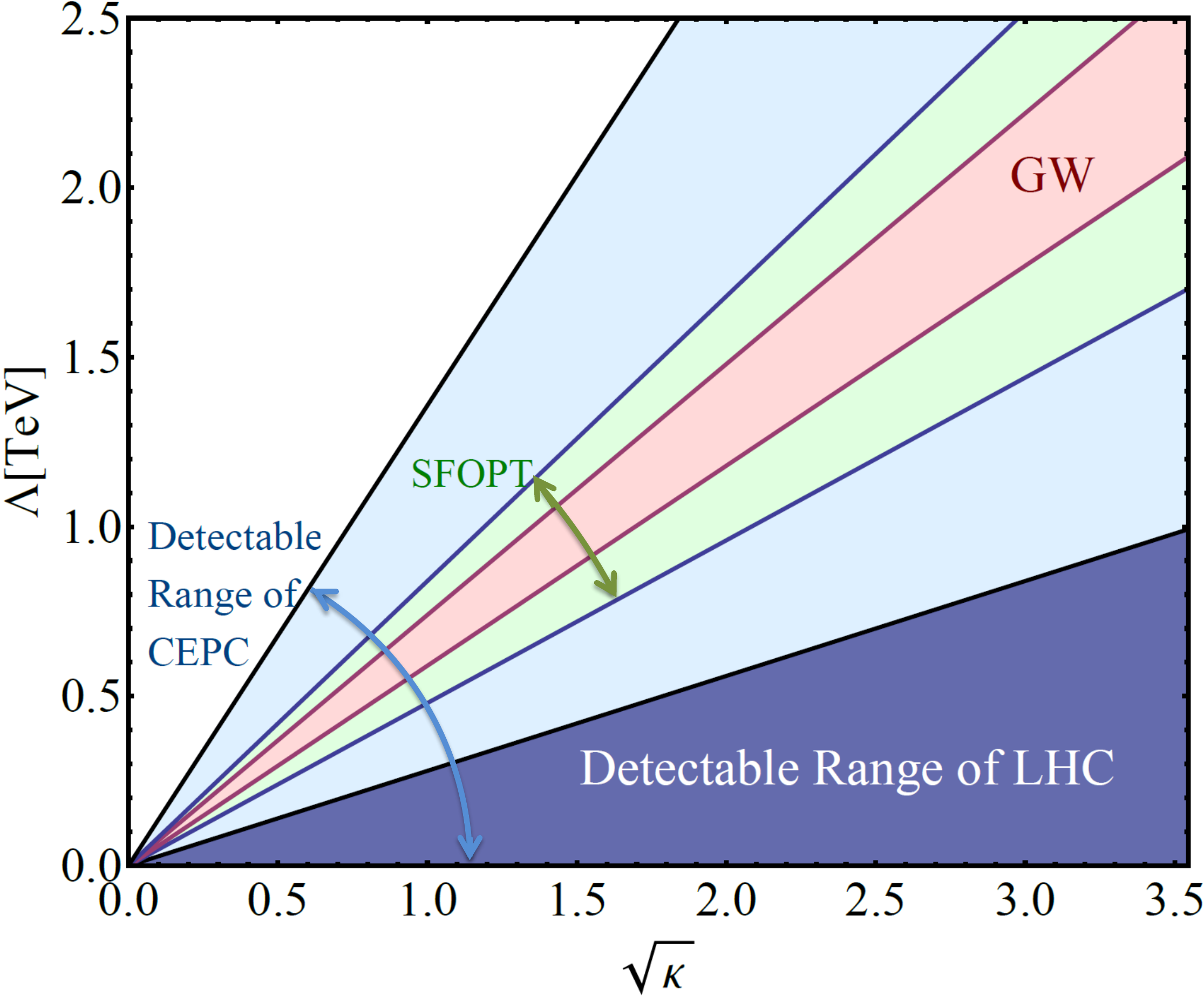}
\caption{The observational abilities of different experiments. For CEPC, the sensitive region is  $\Lambda/\sqrt{\kappa} < 1357.65~{\rm GeV}$; for LHC, it corresponds to $\Lambda/\sqrt{\kappa} < 280~{\rm GeV}$; the theoretical condition for the SFOPT requires $480~ {\rm GeV} < \Lambda/\sqrt{\kappa} < 840~{\rm GeV}$; and the detectable region of GW interferometers reads $590 ~{\rm GeV} < \Lambda/\sqrt{\kappa} < 740 ~{\rm GeV}$. }
\label{pa}
\end{center}
\end{figure}

{\it Results and Discussions.---}
In Fig.~\ref{gw}, the GW spectra $h^2\Omega_{GW}$ and the $hZ$ cross section deviations $\delta_{\sigma_{hZ}}$ are presented by taking different values of the cutoff scale $\Lambda$ (590~GeV, 600~GeV, 650~GeV, 700 GeV and~750~GeV) with $\kappa$ being fixed to unity in the Higgs scenario under consideration. For instance, the red curve in the figure depicts the GW intensity for $\Lambda = 590~{\rm GeV}$ predicted by our model, which also predicts a collider signature of the cross section deviation $\delta_{\sigma_{hZ}} \simeq 2.2\%$ (the corresponding deviation of the trilinear Higgs coupling $\delta_h$ is 1.32), and hence, is expected to be tested at the CEPC. In addition, we numerically present the theoretical curves for the cases of 600 GeV, 650 GeV, 700 GeV and 750 GeV, as shown by the magenta, blue, purple and black lines, respectively. These curves correspond respectively to the values of $2.1\%$, $1.8\%$, and $1.5\%$ for $\delta_{\sigma_{hZ}}$.

From our result, it is obvious that the amplitude of the GW spectrum is more significant for smaller cutoff scales. This fact can be naturally explained by the observation that in Eq.~(\ref{v0}) a smaller $\Lambda$ yields a larger contribution of the sextic operator which then leads to a stronger EWPT. Moreover, it can be found that the GW signals are peaked in the region of $(10^{-4}, 1)$~Hz, which lies in the detectable range of satellite based GW experiments. The colored regions in Fig.~\ref{gw} show the expected experimental sensitivities of various GW interferometers including aLIGO, eLISA\footnote{The eLISA C1 and C4 in the figure are two representative configurations studied in Ref.~\cite{Caprini:2015zlo}.}~\cite{Caprini:2015zlo}, BBO, DECIGO~\cite{Moore:2014lga}, and Ultimate-DECIGO (U-DECIGO)~\cite{Kudoh:2005as}. From Fig.~\ref{gw}, one can explicitly see that eLISA, BBO and U-DECIGO are capable of detecting the GW spectra from the EWPT at low cutoff scales in our model.

{\it Conclusion---}
Consequently, the colliders in particle physics and the GW surveys are naturally correlated through the nature of the EWPT. 
As shown in Fig.~\ref{gw}, each line relates the GW spectrum to the associated collider signal with the same cutoff scale. To obtain a global picture of this correlation, we numerically estimate the observational abilities of different experiments in Fig.~\ref{pa}. For the CEPC with $\sqrt{s}=240~\rm GeV$, the sensitive region is $\Lambda/\sqrt{\kappa} < 1357.65~{\rm GeV}$; for LHC, it corresponds to $\Lambda/\sqrt{\kappa} < 280~{\rm GeV}$; the theoretical condition for the SFOPT roughly requires $480~ {\rm GeV} < \Lambda/\sqrt{\kappa} < 840~{\rm GeV}$; and the detectable region of GW interferometers reads $590 ~{\rm GeV} < \Lambda/\sqrt{\kappa} < 740 ~{\rm GeV}$. From Fig.~\ref{pa}, we find that, to probe the EWPT, the detectable ability of the LHC is relatively weak, but the CEPC and GW detectors are very promising in precisely detecting or even measuring the predicted signals. For example, for the cutoff scale $\Lambda = 590~{\rm GeV}$, the deviation of the trilinear Higgs coupling is $1.32$. While this deviation can not be tested at the LHC, the GW experiments may indirectly measure it, which corresponds to the red line in Fig.~\ref{gw}. We conclude that the GW interferometers can provide a complementary approach to probe the nature of the EWPT alternative to particle colliders, and vice versa.

The recently announced aLIGO observation has initiated a new era of exploring fundamental physics
\cite{Calabrese:2016bnu}. Moreover, after the discovery of the $125$GeV Higgs boson, it becomes urgent to unravel the nature of the EWPT. If this transition was a strong first-order process, it could naturally relate the EW baryogenesis to the GW physics. We present a joint investigation of the possibly observable signatures of this process from both the particle colliders and the GW experiments. Our results show that the GW spectrum produced from the EWPT can be significant enough to be detected by the forthcoming GW experiments. Note that, it is interesting to take into account even higher order operators beyond dimension-6 ones, which, although not very sensitive to be tested in colliders, could be sensitive in GW surveys since the GW signals strongly rely on the details of EWPT. This could further demonstrate the importance of the joint analysis of GW astronomy and particle physics.

The analysis reported in this Letter will contribute to deeply understand the physics of EW baryogenesis, which can build an innovative connection between astrophysics and particle physics. Our joint study will enable novel insights into the astrophysics, GW physics and fundamental particle physics.
Moreover, we have made use of the effective field theory in our study, of which the approach is applied in various disciplines of applied science, such as materials science, condensed matter physics, and physical chemistry. We expect that our results of EWPT and its GW signatures may offer a fresh viewpoint on phase changes and the kinetics of materials in applied science, and reversely, the experimental simulation technology in applied science may shed light on the search of GW/collider signals arisen from the EWPT. Therefore, out study could stimulate the immediate interest of researchers in a broad range of the aforementioned disciplines.

\textit{Acknowledgements.---}
We thank Chiara Caprini, Germano Nardini and Teruaki Suyama for valuable comments.
FPH, YPW and XZ are supported in part by the NSFC (Grant Nos. 11121092, 11033005, 11375220) and by the CAS pilotB program.
FPH is also supported by the China Postdoctoral Science Foundation under Grant No. 2016M590133.
DGW and YFC are supported in part by the Chinese National Youth Thousand Talents Program, by the USTC start-up funding (Grant No. KY2030000049) and by the NSFC (Grant No. 11421303).
The operation of the super-computation is funded by the particle cosmology group at USTC.


\begin{thebibliography}{999}

\bibitem{Abbott:2016blz}
  B.~P.~Abbott {\it et al.} [LIGO Scientific and Virgo Collaborations],
  Phys.\ Rev.\ Lett.\  {\bf 116}, no. 6, 061102 (2016)
  [arXiv:1602.03837 [gr-qc]].

\bibitem{Witten:1984rs}
  E.~Witten,
  Phys.\ Rev.\ D {\bf 30}, 272 (1984).

\bibitem{Hogan:1984hx}
  C.~J.~Hogan,
  Phys.\ Lett.\ B {\bf 133}, 172 (1983);
%
  C.~J.~Hogan,
  Mon.\ Not.\ Roy.\ Astron.\ Soc.\  {\bf 218}, 629 (1986).

\bibitem{Turner:1990rc}
  M.~S.~Turner and F.~Wilczek,
  Phys.\ Rev.\ Lett.\  {\bf 65}, 3080 (1990).

\bibitem{Schwaller:2015tja}
  P.~Schwaller,
  Phys.\ Rev.\ Lett.\  {\bf 115}, 181101 (2015)
  [arXiv:1504.07263 [hep-ph]].

\bibitem{Aad:2012tfa}
  G.~Aad {\it et al.} [ATLAS Collaboration],
  Phys.\ Lett.\ B {\bf 716}, 1 (2012)
  [arXiv:1207.7214 [hep-ex]];
%
  S.~Chatrchyan {\it et al.} [CMS Collaboration],
  Phys.\ Lett.\ B {\bf 716}, 30 (2012)
  [arXiv:1207.7235 [hep-ex]].

\bibitem{CEPC-SPPCStudyGroup:2015csa}
  CEPC-SPPC Study Group,
  IHEP-CEPC-DR-2015-01, IHEP-TH-2015-01, HEP-EP-2015-01.

\bibitem{Arkani-Hamed:2015vfh}
  N.~Arkani-Hamed, T.~Han, M.~Mangano and L.~T.~Wang,
  arXiv:1511.06495 [hep-ph].

\bibitem{Grojean:2006bp}
  C.~Grojean and G.~Servant,
  Phys.\ Rev.\ D {\bf 75}, 043507 (2007)
  [hep-ph/0607107];

\bibitem{Delaunay:2007wb}
  C.~Delaunay, C.~Grojean and J.~D.~Wells,
  JHEP {\bf 0804}, 029 (2008)
  [arXiv:0711.2511 [hep-ph]].

\bibitem{Huber:2007vva}
  S.~J.~Huber and T.~Konstandin,
  JCAP {\bf 0805}, 017 (2008)
  [arXiv:0709.2091 [hep-ph]].

\bibitem{Das:2009ue}
  S.~Das, P.~J.~Fox, A.~Kumar and N.~Weiner,
  JHEP {\bf 1011}, 108 (2010)
  [arXiv:0910.1262 [hep-ph]].

\bibitem{Espinosa:2008kw}
  J.~R.~Espinosa, T.~Konstandin, J.~M.~No and M.~Quiros,
  Phys.\ Rev.\ D {\bf 78}, 123528 (2008)
  [arXiv:0809.3215 [hep-ph]].

\bibitem{Jarvinen:2009mh}
  M.~Jarvinen, C.~Kouvaris and F.~Sannino,
  Phys.\ Rev.\ D {\bf 81}, 064027 (2010)
  [arXiv:0911.4096 [hep-ph]].

\bibitem{Kakizaki:2015wua}
  M.~Kakizaki, S.~Kanemura and T.~Matsui,
  Phys.\ Rev.\ D {\bf 92}, 115007 (2015)
  [arXiv:1509.08394 [hep-ph]].

\bibitem{Zhang:1992fs}
  X.~M.~Zhang,
  Phys.\ Rev.\ D {\bf 47}, 3065 (1993)
  [hep-ph/9301277].

\bibitem{Zhang:1993vh}
  X.~Zhang and B.~L.~Young,
  Phys.\ Rev.\ D {\bf 49}, 563 (1994)
  [hep-ph/9309269];
%
  X.~Zhang, B.~L.~Young and S.~K.~Lee,
  Phys.\ Rev.\ D {\bf 51}, 5327 (1995)
  [hep-ph/9406322];
%
  X.~Zhang, S.~K.~Lee, K.~Whisnant and B.~L.~Young,
  Phys.\ Rev.\ D {\bf 50}, 7042 (1994)
  [hep-ph/9407259];
%
  K.~Whisnant, B.~L.~Young and X.~Zhang,
  Phys.\ Rev.\ D {\bf 52}, 3115 (1995)
  [hep-ph/9410369].

\bibitem{Grzadkowski:2003sd}
  B.~Grzadkowski, J.~Pliszka and J.~Wudka,
  Phys.\ Rev.\ D {\bf 69}, 033001 (2004)
  [hep-ph/0307338].

\bibitem{Huang:2015bta}
  F.~P.~Huang and C.~S.~Li,
  Phys.\ Rev.\ D {\bf 92}, 075014 (2015)
  [arXiv:1507.08168 [hep-ph]].

  
\bibitem{Huang:2015izx}
  F.~P.~Huang, P.~H.~Gu, P.~F.~Yin, Z.~H.~Yu and X.~Zhang,
  Phys.\ Rev.\ D {\bf 93} (2016) no.10,  103515
  doi:10.1103/PhysRevD.93.103515
  [arXiv:1511.03969 [hep-ph]].

\bibitem{TheLIGOScientific:2014jea}
  J.~Aasi {\it et al.} [LIGO Scientific Collaboration],
  Class.\ Quant.\ Grav.\  {\bf 32}, 074001 (2015)
  doi:10.1088/0264-9381/32/7/074001
  [arXiv:1411.4547 [gr-qc]].

\bibitem{TheVirgo:2014hva}
  F.~Acernese {\it et al.} [VIRGO Collaboration],
  Class.\ Quant.\ Grav.\  {\bf 32}, no. 2, 024001 (2015)
  doi:10.1088/0264-9381/32/2/024001
  [arXiv:1408.3978 [gr-qc]].

\bibitem{Seoane:2013qna}
  P.~A.~Seoane {\it et al.} [eLISA Collaboration],
  arXiv:1305.5720 [astro-ph.CO].

\bibitem{Seto:2001qf}
  N.~Seto, S.~Kawamura and T.~Nakamura,
  Phys.\ Rev.\ Lett.\  {\bf 87}, 221103 (2001)
  [astro-ph/0108011];
%
  S.~Kawamura {\it et al.},
  Class.\ Quant.\ Grav.\  {\bf 23}, S125 (2006);
%
  S.~Kawamura {\it et al.},
  Class.\ Quant.\ Grav.\  {\bf 28}, 094011 (2011).

\bibitem{Corbin:2005ny}
  V.~Corbin and N.~J.~Cornish,
  Class.\ Quant.\ Grav.\  {\bf 23}, 2435 (2006)
  [gr-qc/0512039].

\bibitem{Quiros:1999jp}
  M.~Quiros,
  hep-ph/9901312.

\bibitem{Gomez-Ceballos:2013zzn}
  M.~Bicer {\it et al.} [TLEP Design Study Working Group Collaboration],
  JHEP {\bf 1401}, 164 (2014)
  [arXiv:1308.6176 [hep-ex]].

\bibitem{Grojean:2004xa}
  C.~Grojean, G.~Servant and J.~D.~Wells,
  Phys.\ Rev.\ D {\bf 71}, 036001 (2005)
  [hep-ph/0407019].

\bibitem{Noble:2007kk}
  A.~Noble and M.~Perelstein,
  Phys.\ Rev.\ D {\bf 78}, 063518 (2008)
  [arXiv:0711.3018 [hep-ph]].

\bibitem{Kamionkowski:1993fg}
  M.~Kamionkowski, A.~Kosowsky and M.~S.~Turner,
  Phys.\ Rev.\ D {\bf 49}, 2837 (1994)
  [astro-ph/9310044].

\bibitem{Hindmarsh:2013xza}
  M.~Hindmarsh, S.~J.~Huber, K.~Rummukainen and D.~J.~Weir,
  Phys.\ Rev.\ Lett.\  {\bf 112}, 041301 (2014)
  [arXiv:1304.2433 [hep-ph]];
%
  M.~Hindmarsh, S.~J.~Huber, K.~Rummukainen and D.~J.~Weir,
  Phys.\ Rev.\ D {\bf 92}, 123009 (2015)
  [arXiv:1504.03291 [astro-ph.CO]].

\bibitem{Kosowsky:2001xp}
  A.~Kosowsky, A.~Mack and T.~Kahniashvili,
  Phys.\ Rev.\ D {\bf 66}, 024030 (2002)
  [astro-ph/0111483].




\bibitem{Caprini:2009yp}
  C.~Caprini, R.~Durrer and G.~Servant,
  JCAP {\bf 0912}, 024 (2009)
  [arXiv:0909.0622 [astro-ph.CO]].

\bibitem{Huber:2008hg}
  S.~J.~Huber and T.~Konstandin,
  JCAP {\bf 0809}, 022 (2008)
  [arXiv:0806.1828 [hep-ph]].

\bibitem{Caprini:2015zlo}
  C.~Caprini {\it et al.},
  arXiv:1512.06239 [astro-ph.CO].

\bibitem{Espinosa:2010hh}
  J.~R.~Espinosa, T.~Konstandin, J.~M.~No and G.~Servant,
  JCAP {\bf 1006}, 028 (2010)
  [arXiv:1004.4187 [hep-ph]].

\bibitem{Binetruy:2012ze}
  P.~Binetruy, A.~Bohe, C.~Caprini and J.~F.~Dufaux,
  JCAP {\bf 1206}, 027 (2012)
  [arXiv:1201.0983 [gr-qc]].

\bibitem{Bodeker:2009qy}
  D.~Bodeker and G.~D.~Moore,
  JCAP {\bf 0905}, 009 (2009)
  [arXiv:0903.4099 [hep-ph]].

\bibitem{Huber:2013kj}
  S.~J.~Huber and M.~Sopena,
  arXiv:1302.1044 [hep-ph].



\bibitem{Moore:2014lga}
  C.~J.~Moore, R.~H.~Cole and C.~P.~L.~Berry,
  Class.\ Quant.\ Grav.\  {\bf 32}, 015014 (2015)
  [arXiv:1408.0740 [gr-qc]].

\bibitem{Kudoh:2005as}
  H.~Kudoh, A.~Taruya, T.~Hiramatsu and Y.~Himemoto,
  Phys.\ Rev.\ D {\bf 73}, 064006 (2006)
  [gr-qc/0511145].

\bibitem{Calabrese:2016bnu}
  E.~Calabrese, N.~Battaglia and D.~N.~Spergel,
  arXiv:1602.03883 [gr-qc].

\end{thebibliography}
\end{document}